\documentclass[
aps,
prx,
superscriptaddress,
amsmath, amssymb,
amssymb,
reprint,
floatfix,
longbibliography
]{revtex4-1}

\usepackage{graphicx}
\usepackage{bm}
\usepackage{color}
\usepackage[]{natbib}
\usepackage{amsmath,amssymb,amsfonts}
\usepackage[normalem]{ulem}
\usepackage{hyperref}	

\definecolor{pink}{rgb}{0.858, 0.188, 0.478}

\begin{document}

\title{Large dispersive interaction between a CMOS double quantum dot and microwave photons}

\author{David J. Ibberson}\email{david.ibberson@bristol.ac.uk}
\affiliation{Quantum Engineering Technology Labs, University of Bristol, Tyndall Avenue, Bristol BS8 1FD, U.K.}
\affiliation{Hitachi Cambridge Laboratory, J.J. Thomson Avenue, Cambridge CB3 0HE, U.K.}
\affiliation{Quantum Engineering Centre for Doctoral Training, University of Bristol, Tyndall Avenue, Bristol BS8 1FD, U.K.}
\author{Theodor Lundberg}
\affiliation{Cavendish Laboratory, University of Cambridge, J.J. Thomson Avenue, Cambridge CB3 0HE, U.K.}
\affiliation{Hitachi Cambridge Laboratory, J.J. Thomson Avenue, Cambridge CB3 0HE, U.K.}
\author{James A. Haigh}
\affiliation{Hitachi Cambridge Laboratory, J.J. Thomson Avenue, Cambridge CB3 0HE, U.K.}
\author{Louis Hutin}
\affiliation{CEA/LETI-MINATEC, CEA-Grenoble, 38000 Grenoble, France}
\author{Benoit Bertrand}
\affiliation{CEA/LETI-MINATEC, CEA-Grenoble, 38000 Grenoble, France}
\author{Sylvain Barraud}
\affiliation{CEA/LETI-MINATEC, CEA-Grenoble, 38000 Grenoble, France}
\author{Chang-Min Lee}
\affiliation{Department of Materials Science \& Metallurgy, University of Cambridge, 27 Charles Babbage Road, Cambridge CB3 0FS, U.K.}
\author{Nadia A. Stelmashenko}
\affiliation{Department of Materials Science \& Metallurgy, University of Cambridge, 27 Charles Babbage Road, Cambridge CB3 0FS, U.K.}
\author{Giovanni A. Oakes}
\affiliation{Cavendish Laboratory, University of Cambridge, J.J. Thomson Avenue, Cambridge CB3 0HE, U.K.}
\affiliation{Hitachi Cambridge Laboratory, J.J. Thomson Avenue, Cambridge CB3 0HE, U.K.}
\author{Laurence Cochrane}
\affiliation{Nanoscience Centre, Department of Engineering, University of Cambridge, Cambridge CB3 0FF, U.K.}
\affiliation{Hitachi Cambridge Laboratory, J.J. Thomson Avenue, Cambridge CB3 0HE, U.K.}
\author{Jason W. A. Robinson}
\affiliation{Department of Materials Science \& Metallurgy, University of Cambridge, 27 Charles Babbage Road, Cambridge
CB3 0FS, U.K.}
\author{Maud Vinet}
\affiliation{CEA/LETI-MINATEC, CEA-Grenoble, 38000 Grenoble, France}
\author{M. Fernando Gonzalez-Zalba}\email{mg507@cam.ac.uk}\altaffiliation[Current address: ]{Quantum Motion Technologies, Windsor House, Cornwall Road, Harrogate HG1 2PW, U.K.}
\affiliation{Hitachi Cambridge Laboratory, J.J. Thomson Avenue, Cambridge CB3 0HE, U.K.}
\author{Lisa A. Ibberson}
\affiliation{Hitachi Cambridge Laboratory, J.J. Thomson Avenue, Cambridge CB3 0HE, U.K.}

\begin{abstract}
We report fast charge state readout of a double quantum dot in a CMOS split-gate silicon nanowire transistor via the large dispersive interaction with microwave photons in a lumped-element resonator formed by hybrid integration with a superconducting inductor. We achieve a coupling rate $g_0/(2\pi) = 204 \pm 2$~MHz by exploiting the large interdot gate lever arm of an asymmetric split-gate device, $\alpha=0.72$, and by inductively coupling to the resonator to increase its impedance, $Z_\text{r}=560$~$\Omega$. In the dispersive regime, the large coupling strength at the double quantum dot hybridisation point produces a frequency shift comparable to the resonator linewidth, the optimal setting for maximum state visibility. We exploit this regime to demonstrate rapid dispersive readout of the charge degree of freedom, with a SNR of 3.3 in 50~ns. In the resonant regime, the fast charge decoherence rate precludes reaching the strong coupling regime, but we show a clear route to spin-photon circuit quantum electrodynamics using hybrid CMOS systems.
\end{abstract}

\maketitle

\section{Introduction}

Electron spins in silicon quantum dots are promising candidates for large-scale quantum computation~\cite{Zwanenburg2013}. Their long coherence times enable high-fidelity single- and two- qubit gates~\cite{yoneda2018quantum,Yang2019,Huang2019, Xue2019, zajac2018resonantly} and their compatibility with industrial manufacturing processes ~\cite{hutin2018si, pillarisetty2019high, zwerver2021qubits} facilitates the fabrication of large quantum dot arrays, and their integration with classical electronics~\cite{schaal2019cmos}. All quantum computing architectures aiming to achieve fault tolerance~\cite{Fowler2012} require high-fidelity quantum state readout. To implement quantum error correction (QEC), this readout must be achieved in time scales significantly shorter than the coherence time of the qubits ($T_2 = 28$~ms \cite{veldhorst2014addressable}), and ideally faster than the time scales of single- and two-qubit gates to avoid becoming the bottleneck in error-correction cycles. For the highest fidelity gates, these timescales are sub-$\mu$s~\cite{Yang2019} and 5~$\mu$s \cite{Huang2019}, respectively. Renewed efforts have been directed at achieving this goal using charge sensors combined with amplification at the millikelvin stage~\cite{Curry2019} or with radio-frequency techniques~\cite{Urdampilleta2019, volk2019fast}, reaching a promising 97\% fidelity in $1.5~\mu$s \cite{Keith2019}.

Circuit quantum electrodynamics (cQED) provides an alternative readout approach which removes the necessity for a charge sensor and simplifies the qubit architecture \cite{Blais2004, wallraff2004strong, blais2020circuit}. The qubit interacts with photons in a detuned microwave cavity whose eigenfrequencies are modified conditional on the qubit state. Recent results have demonstrated this is a competitive approach for silicon spin qubits and have reached fidelities of 73\% in 2.6 ms~\cite{west2019gate} and 82.9\% in 300 $\mu$s~\cite{Pakkiam2018}, when using hybrid resonant circuits, and 98\% in 6 $\mu$s, when using on-chip resonators~\cite{zheng2019rapid}. Although on-chip integration has enabled faster readout, it poses an important challenge for the future development of compact nanometre-scale spin qubit arrays, given that current microwave resonator designs have characteristic length scales of the order of $100~\mu$m~\cite{samkharadze2018strong, stockklauser2017strong}. Hybrid multi-chip modules give freedom to optimise the layout of the qubit layer without the spatial constraints of the microwave electronics and are likely to facilitate the development of two-dimensional qubit arrays \cite{Veldhorst2017, Li2018}, a requirement for the implementation of the surface code QEC~\cite{Fowler2012}. However, the technology to achieve readout times compatible with QEC using hybrid multi-chip modules requires further development.

In this paper, we demonstrate fast charge state readout of a complementary-metal-oxide-semiconductor (CMOS) double quantum dot (DQD) coupled to a high-impedance lumped-element resonator. We measure an interdot charge transition with a signal-to-noise ratio ($\text{SNR}$) of 3.3 in just 50 ns. The signal, which represents that of a singlet outcome in singlet-triplet readout schemes, should enable high-fidelity spin readout in sub-$\mu$s time scales. Our large signal is achieved by increasing the charge-photon coherent coupling rate to $g_0/(2\pi)=204$~MHz, comparable to those reported in fully-integrated circuits~\cite{stockklauser2017strong, samkharadze2018strong}. The large coupling results directly from exploiting the large interdot gate lever arm of an asymmetric split-gate device, $\alpha=0.72$, and from inductively coupling to the resonator to increase its impedance. It allows us to operate the system in the intermediate dispersive regime, where the dispersive frequency shift of the microwave cavity, $\chi$, is as large as the photon decay rate, $\kappa$; the condition for maximum state visibility~\cite{schuster2007thesis}. Furthermore, we show the scalability of the architecture by simultaneously reading two resonators, each of them containing a quantum dot device, via frequency multiplexing. Finally, by tuning the charge configuration of the DQD, we reach the resonant regime of cQED, where the qubit and resonator energies match. Although $g_0\gg\kappa$, the fast decay rate of the charge degree of freedom of the DQD precludes reaching the strong coupling limit in this experiment. By coupling to the long-lived spin degree of freedom \cite{ciriano2020spin}, it may be possible to reach the strong coupling limit, which has been achieved in Si/SiGe heterostructures \cite{samkharadze2018strong, mi2018coherent}, but not yet in Si/SiO$_2$ systems. We note that the microwave circuitry presented here should readily improve the readout fidelity not only of CMOS-based devices but of any semiconductor-based system, especially those with a large gate lever arm such as topological qubits~\cite{de2019rapid, sabonis2019dispersive}.

\section{Architecture}

\begin{figure*}
\centering
\includegraphics[scale=1]{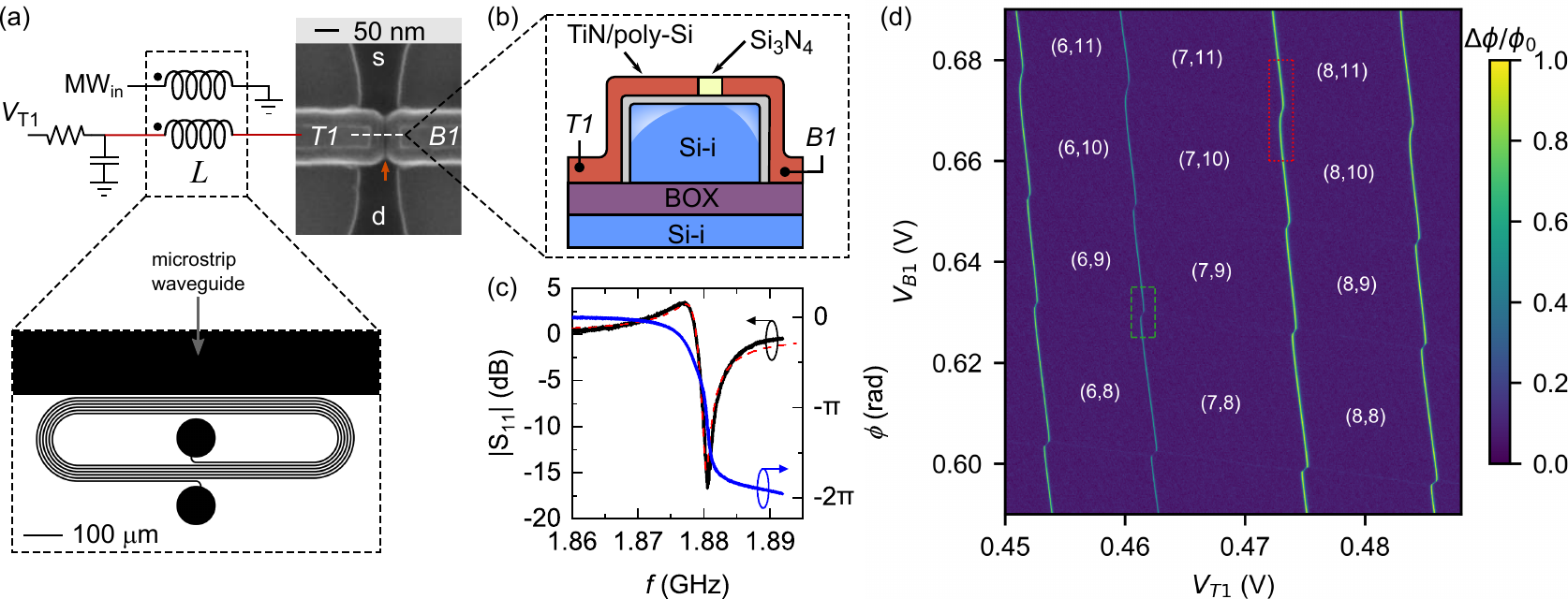}
\caption{\textbf{Measurement set-up, resonator geometry and double quantum dot characteristics.} (a) Two mutually coupled inductors represent the inductive coupling between a microstrip waveguide and a spiral inductor with inductance $L$, the layout of which is shown below in the dashed box. Both elements are fabricated from the same NbN film on sapphire; the gap separating inductor and waveguide is 4~$\mu$m wide. The inductor is wire-bonded to a NWFET on a separate substrate (aluminium wire bonds are highlighted in red). The bias-tee consists of a 100 pF capacitor and a 100~k$\Omega$ resistor. An SEM image shows a top-down view of a similar split-gate NWFET, where the orange arrow highlights the insulating spacer between the two top gates and its misalignment from the centre of the nanowire. (b) Cross-section schematic of the NWFET, cutting through the nanowire perpendicular to its length. Quantum dots are formed in the top corners of the intrinsic silicon body, indicated by white shading. (c) The phase (blue), amplitude (black) and fit (red) of a reflected microwave signal applied and measured at MW$_\text{in}$ when the terminals of the NWFET are grounded. (d) The DQD charge stability diagram shown in normalized reflected phase data as $V\textsubscript{T1}$ and $V\textsubscript{B1}$ are varied. Numbers in parentheses indicate the electron numbers on each dot; the highlighted ICTs are measured in detail later.}
\label{fig1}
\end{figure*}

Our multi-module assembly consists of a silicon chiplet and a superconducting chiplet, connected via wire-bonds. The superconducting chiplet contains an elongated spiral inductor and a 50~$\Omega$ microstrip waveguide, fabricated using optical lithography from an 80~nm thick sputter-deposited NbN film on a sapphire substrate, shown in Fig.~\ref{fig1}(a). The structure is designed to allow microwave radiation to be inductively coupled into the inductor from the waveguide. We connect the inner bond pad of the inductor to the T1 gate of an $n$-type split-gate silicon-on-insulator nanowire field-effect transistor (NWFET), see Fig.~\ref{fig1}(b), and the outer bond pad to a bias-tee to enable DC gating. More details of the circuit can be found in Appendix \ref{appendix:circuit}.

The intrinsic-Si body of the nanowire, with a height of 7nm and width 70~nm, is covered by 6~nm SiO$_2$ gate oxide and separated from the silicon substrate by a 145~nm buried oxide. Two gates, arranged face-to-face, wrap around the sides and part of the top surface of the nanowire. The gates have a length along the nanowire of 60~nm, and surrounding the gates the nanowire is covered by 34~nm Si$_3$N$_4$ spacers. The gap between the gates at the top of the nanowire is 40~nm wide, and is also covered by Si$_3$N$_4$~\cite{lundberg2019spin}. An orange arrow on the top-down SEM image, Fig.~\ref{fig1}(a), highlights the gap between the gates. We connect the inductor to the gate with the larger overlap to make use of the larger interdot gate lever arm, as we will see later. Under the appropriate bias, a DQD forms within the corners of the NWFET, as illustrated in Fig.~\ref{fig1}(b)~\cite{betz2015dispersively}. We model a single valence electron in the DQD as a coupled two-level system with eigenfrequencies $\Omega_\pm=\pm\frac{1}{2}\Omega=\pm\frac{1}{2}\sqrt{\varepsilon^2+(2t_\text{c})^2}$ where $t_\text{c}$ is the tunnel coupling and $\varepsilon$ is the frequency detuning between dots.

The combined inductor-DQD system forms an $LC$ resonant circuit. We study the properties of the resonator by monitoring the amplitude and phase of the reflected signal when applying a microwave tone to the microwave input line, MW$_\text{in}$, with the NWFET grounded at 12~mK, see Fig.~\ref{fig1}(c). We find a resonance at $f_\text{0}$ = 1.88~GHz, which can be fitted using a complex external Q-value (see red dashed line in Fig.~\ref{fig1}(c)) \cite{wisbey2010effect, khalil2012analysis}. From the fit and our experimentally measured inductance value $L=47$~nH \cite{ahmed2018radio}, we deduce a total parasitic capacitance $C_\text{p}=150$~fF, a characteristic impedance $Z_\text{r}=\sqrt{L/C_\text{p}}=560$~$\Omega$, an internal (external) decay rate $\kappa_\text{int}/(2\pi)=1.31 $~MHz ($\kappa_\text{ext}/(2\pi)=1.76$~MHz) and a total photon decay rate of $\kappa/(2\pi)=3.07$~MHz. Moreover, we plot the phase shift $\phi$ across the resonance (blue trace) and observe a $2\pi$ phase shift confirming that photons predominantly escape the resonator before decaying~\cite{ibberson2019low}.

Next, at fixed frequency, we monitor the phase shift of the reflected signal $\phi$ as a function of the gate biases $V_\text{T1}$ and $V_\text{B1}$, see Fig.~\ref{fig1}(d). An enhancement of the phase signal in a quasi-periodic honeycomb pattern is clearly visible. The lines correspond to charge bistable regions in a DQD where the charge susceptibility of the DQD is non-zero~\cite{mizuta2017quantum, benito2017input}. We infer the electron population in each quantum dot by using a visible dot-to-reservoir transition of one dot as a charge sensor of the other dot, see Appendix \ref{appendix:elecronpop}. Then we invert the role of the dots and count the number of electrons in the opposite dot~\cite{lundberg2019spin}. The numbers in parentheses in Fig. \ref{fig1}(d) indicate the number of electrons on the dot under the (T1, B1) gates. Interdot charge transitions (ICTs) at intersections of the horizontal and vertical lines are not visible for electron populations smaller than those presented on this plot, indicating insufficient wavefunction overlap for tunneling to occur in the timescale of the resonator period \cite{gilbert2020single}.

In the remainder of this paper, we focus on studying the interaction between the resonator and the DQD. We use two ICTs with different tunnel couplings to access two different regimes. First, the dispersive regime, where the qubit frequency is greater than the resonator frequency, which we use to achieve fast readout of the charge state (red dotted box in Fig. \ref{fig1}(d)) and second, the resonant regime, where the qubit and resonator frequencies are equal (green dashed box in Fig. \ref{fig1}(d)).

\begin{figure*}
\centering
\includegraphics[scale=1]{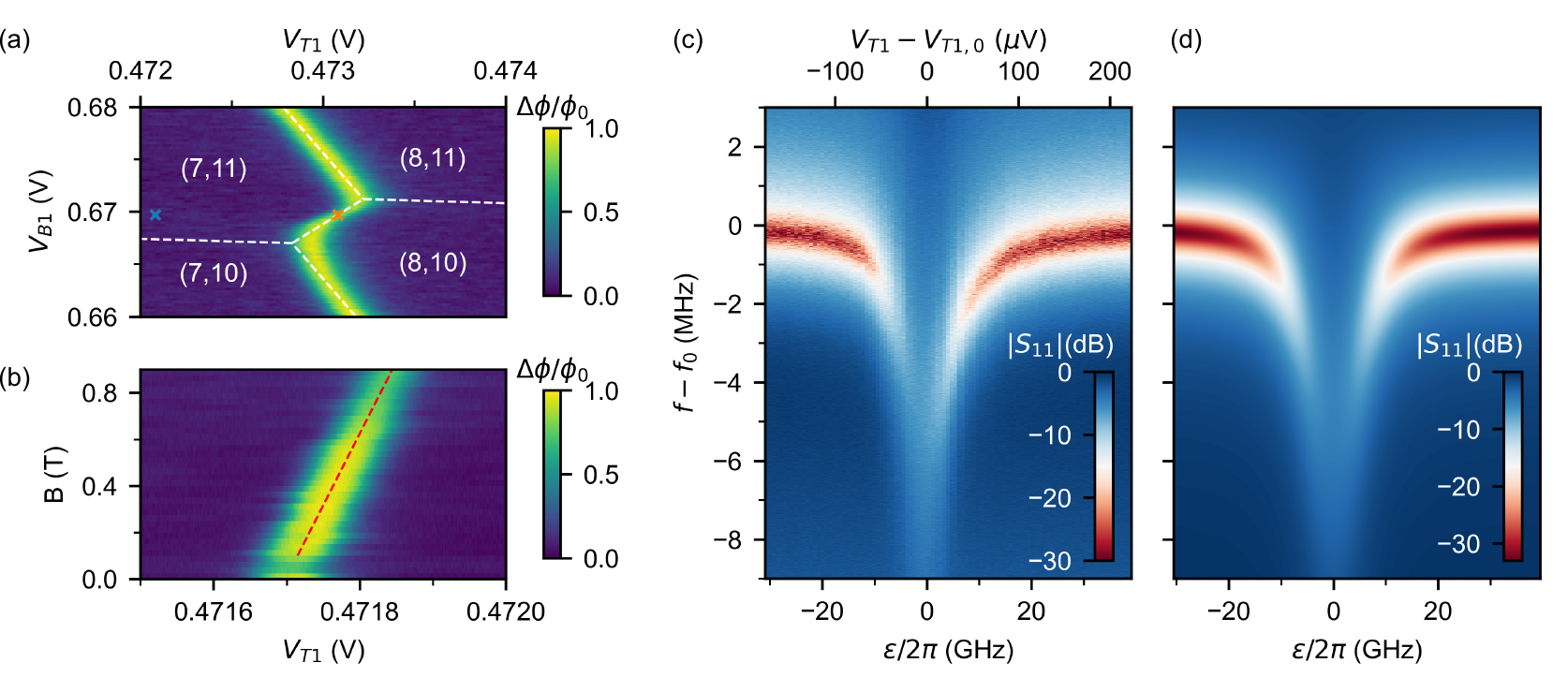}
\caption{\textbf{Resonator-DQD interaction in the dispersive regime.} (a) Reduced phase response of the resonator $\Delta \phi/\phi_0$ around the $(7,11)\longleftrightarrow (8,10)$ ICT. Colour-coded crosses denote the positions of measurements on and off the ICT respectively, see Fig.~\ref{fig3}(a). (b) Magnetospectroscopy of the $(7,11)\longleftrightarrow (8,10)$ ICT, which we used to calculate the interdot gate lever arm. (c) Magnitude of the signal reflected from the cavity showing the shift in the resonant frequency as $V_\text{T1}$ is swept across the ICT. (d) Simulated data, using Eq.~\ref{eq1}, to reproduce the experimental data in panel (c).}
\label{fig2}
\end{figure*}

\section{Large dispersive interaction}

We bias the DQD to the ICT between charge states (7,11) and (8,10), see  Fig.~\ref{fig2}(a). The ICT and the T1 dot-to-reservoir (DTR) lines are visible, but the B1 DTR ones are not because of the weak lever arm of the T1 gate to the B1 dot. We focus on the ICT and extract the interdot gate lever arm from the voltage shift of the transition as the magnetic field is increased, see Fig.~\ref{fig2}(b). For an even parity transition such as this, where the total number of valence electrons is even (for example a $(1,1) \leftrightarrow (2,0)$ transition), the position of the ICT shifts with a slope determined by $\alpha/g_\text{e}\mu_\text{B}$, where $g_\text{e}$ is the electron $g$-factor $\sim 2$ and $\mu_\text{B}$ is the Bohr magneton~\cite{mizuta2017quantum}. The signal at $B > 0$~T originates from charge transitions between states with different spin numbers. The reason for spin transitions occurring under the microwave electric field is unknown, but could potentially be explained by electrically driven Landau-Zener transitions followed by a fast relaxation hotspot near the point of anticrossing \cite{yang2013spin}. We obtain $\alpha=0.724 \pm 0.006$. The large $\alpha$ is a direct consequence of the large asymmetric channel overlap of T1 and represents the highest reported for any silicon device. Previous measurements of a device from the same wafer used the gate electrode with the smaller overlap (B1) which yielded $\alpha=0.345$~\cite{lundberg2019spin}. Achieving a large $\alpha$ is of great importance for cQED architectures since it relates directly to the dipole moment of the quantum system and hence to the coherent coupling constant $g_0/(2\pi)=\alpha f_\text{0}\sqrt{Z_\text{r}/2R_\text{Q}}$, where $R_\text{Q}$ is the resistance quantum~\cite{childress2004mesoscopic}. 

To explore the interaction between the resonator and the DQD, we measure the reflected spectrum of the resonator as we change the DQD frequency detuning, $\varepsilon$, via $V_\text{T1}$, see Fig.~\ref{fig2}(c). We apply the input signal with a power of -125~dBm at $\text{MW}_\text{in}$, well below the onset of power broadening. At $\varepsilon=0$, the resonator exhibits a maximum frequency shift with respect to the bare resonator frequency, $\chi/(2\pi) = 4.1195 \pm 0.0006$~MHz. This value is comparable to the bare cavity linewidth $\kappa/(2\pi)$. In addition, we observe an increase in the effective resonator linewidth, $\kappa^*$, which we attribute to the interaction of the resonator with a fast decaying DQD, as we shall confirm later. 

To extract the relevant parameters of the resonator-DQD system, and to simulate the response, we model the DQD as a coupled two-level system, and the microwaves by a weak periodic field~\cite{petersson2012circuit}. In the rotating-wave approximation, we obtain the effective Hamiltonian 

\begin{equation}
	H_\text{tot}=\hbar\Delta_0a^\dagger a+\frac{\hbar\Delta}{2}\hat{\sigma_\text{z}}+\hbar g_\text{eff}(a\sigma_++a^\dagger\sigma_-)
\end{equation}

\noindent where $\Delta_0=(2\pi)(f_\text{0}-f)$, $\Delta=\Omega-(2\pi) f_\text{0}$, $g_\text{eff}=g_02t_\text{c}/\Omega$, $\hat{\sigma_\text{z}}$ is the inversion operator, $\sigma_{+(-)}$ are the raising and lowering operators of the DQD and $a$ and $a^\dagger$ are the photon annihilation and creation operators. We use the input operator formalism and develop the Heisenberg-Langevin equations of motion to obtain the reflection coefficient in the steady state when neglecting quantum noise contributions

\begin{equation}\label{eq1}
\left|S_{11}\right| = \left|1+ \frac{i\kappa_\text{ext}}{\Delta_0-\frac{i\kappa}{2} + g_\text{eff}\chi_\text{DQD}} \right|^2
\end{equation}

\noindent where $|S_{11}|$ is the reflected power ratio, $\chi_\text{DQD}=g_\text{eff}/(-\Delta+i\gamma/2)$ is the susceptibility of the DQD and $\gamma$ is the DQD decay rate. We use Eq.~\ref{eq1} to simulate the experimental results, see Fig.~\ref{fig2}(d). From the $\varepsilon$-dependent shift in the resonance frequency, $\chi=-g_\text{eff}^2\Delta/\left[(\Delta^2+\gamma^2/4)\right]$, and the resonance linewidth increase $\kappa^*=\kappa+g_\text{eff}^2\gamma/(\Delta^2+\gamma^2/4)$, see Appendix \ref{appendix:sparam}, we determine the relevant system parameters: $g_0/(2\pi)=204 \pm 2$~MHz, $2t_\text{c}/ (2\pi)=7.76 \pm 0.13$~GHz and $\gamma/(2\pi)=9.95 \pm 0.29$~GHz. From these values, we can draw some important conclusions. Firstly, the DQD presents a fast decay rate, in line with reports in the literature for the charge degree of freedom in silicon MOS structures~\cite{dupont2013coherent, gonzalez2016gate, chatterjee2018silicon}. We discard Purcell-enhanced relaxation, $\gamma_\text{p}/(2\pi)=(g_0/\Delta)^2\kappa/(2\pi)=$ 3.7~kHz~\cite{blais2004cavity}, and measurement-induced decoherence due to photon shot noise, $\gamma_\phi/(2\pi)=(\sqrt{2\bar{n}}-1)\kappa/(4\pi)=25$~MHz~\cite{yan2016flux} (where $\bar{n}$ is the average photon number) as the cause of the large $\gamma$, and instead attribute it to charge noise at the Si/SiO$_2$ interface. Secondly, although $\gamma\gg g_0$, we note that the coupling strength is much greater than the resonator linewidth (i.e. $g_0\gg\kappa$) and hence our system satisfies one of the requirements to reach the strong coupling regime. Finally, since $\Delta\gg g_0$ and $\chi/\kappa\approx 1$, our measurements are in the intermediate dispersive limit, relevant for high-fidelity qubit readout.

\section{Dispersive Readout Performance}

\begin{figure}
\centering
\includegraphics[scale=1]{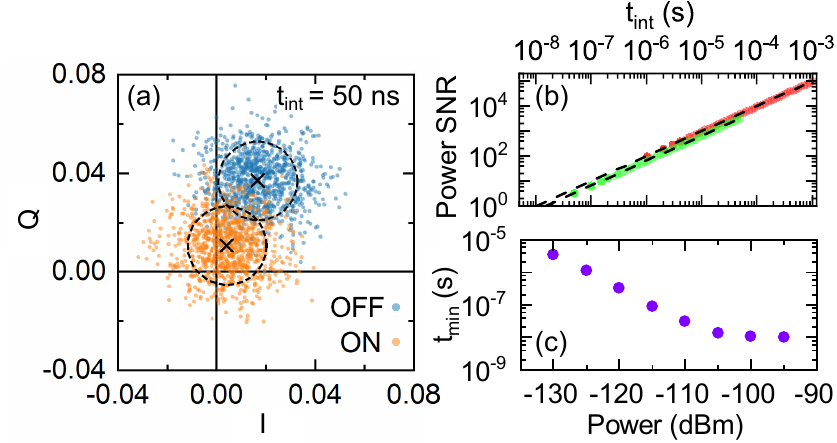}
\caption{\textbf{Readout performance evaluation.} (a) Distribution of the reflected signal in quadrature space, collected at the two points marked in Fig.~\ref{fig2}(a) on and off the ICT signal. Each point was collected with an integration time $t_\text{int}=50$~ns. For each distribution, the black cross marks the mean and the dashed circle indicates the standard deviation of distances to the mean. (b) SNR dependence on $t_\text{int}$ for input power = -100 dBm. Red data points were taken with a 1 MHz low pass filter and the green data points were taken with a 20 MHz low pass filter. The dashed lines extrapolate the data to SNR=1, from which we extract the minimum integration time, $t_{\text{min}}$. (c) Decrease in $t_{\text{min}}$ with increasing input power, which saturates due to power broadening at $\sim 100$~dBm \cite{maman2020charge}.}
\label{fig3}
\end{figure}

We now evaluate the dispersive readout performance by measuring the signal-to-noise ratio (SNR) of the charge transition as a function of integration time, $t_\text{int}$. Using a continuous-wave excitation, we measure the $I$ and $Q$ demodulated components of the reflected signal at the two voltage positions marked in Fig.~\ref{fig2}(a); the Coulomb blockaded region (blue cross) and the charge degenerate regime (orange cross). This corresponds to the expected signal in a singlet-triplet spin readout scheme~\cite{west2019gate}, in the limit where the valley splitting is larger than the tunnel coupling. Measurements of the valley splitting in excess of $500~\mu$eV have been reported for corner dots in the single-electron regime \cite{ciriano2020spin} which is much larger than the tunnel coupling measured in this work, $2t_\text{c} = 32~\mu$eV. As shown in Fig.~\ref{fig3}(a), the distributions of points at the two voltages are separated in the $IQ$ plane. We use this data to calculate the power SNR, given by $(A/B)^2$, where $A$ is the distance between the centers of the clusters (marked by black crosses), and $B$ is the average standard deviation of the two states (dotted circles). In Fig.~\ref{fig3}(b), the red and green dots plot the SNR for two instrument configurations, using either a 1~MHz or 20~MHz maximum measurement bandwidth respectively. From a linear fit to the data, we extract $t_\text{min}$, defined as the integration time for which SNR = 1. At the optimal input power of -100~dBm (see Fig.~\ref{fig3}(c)), for the 1~MHz data we find $t_\text{min}=10$~ns. This value is shorter than other silicon implementations which have benefited from either a Josephson Parametric Amplifier (JPA) ~\cite{schaal2020fast}, or a fully-integrated resonator ~\cite{zheng2019rapid}. Our result is comparable to the best reported in any dispersively detected semiconductor quantum dot, which was obtained in InAs and utilised a JPA~\cite{stehlik2015fast}. Furthermore, following a simple readout fidelity model~\cite{barthel2009rapid}, we obtain a 99.7\% electrical readout fidelity in 300~ns. Using the 20~MHz bandwidth configuration (green dots), we can experimentally access integration times down to 50~ns. We note that the lower SNR in the 20~MHz case is due to the comparably larger equivalent noise bandwidth of the filtering system. We measure an SNR of 3.3 at $t_\text{int}=50$~ns, for which the $IQ$ data is plotted in Fig.~\ref{fig3}(a). This result agrees well with the estimated SNR in the intermediate dispersive regime $\text{SNR} = \left(\Delta^2 hf_0 k_\text{ext} t_\text{int}\right)/\left(8\pi g_0^2 k_\text{B}T_\text{N}\right)=$~2.6, where $k_\text{B}$ is the Boltzmann constant and $T_\text{N}\approx 4$~K is the noise temperature of the cryogenic amplifier, see Appendix~\ref{appendix:readoutheory}.

Finally, we highlight that this architecture can be easily extended to read out several qubits from a single transmission line by adding additional spirals along the length of the microstrip waveguide and employing frequency-multiplexing techniques \cite{hornibrook2014frequency, heinsoo2018rapid}. We simultaneously read the charge states of two separate NWFETs embedded in resonators of different frequencies (1.8 GHz and 2.4 GHz, corresponding to superconducting inductors with $L_1=47$~nH and $L_2=29$~nH respectively), see Appendix \ref{appendix:freqmux} for experimental details. 

\section{Resonant regime}

\begin{figure}
\centering
\includegraphics[scale=1]{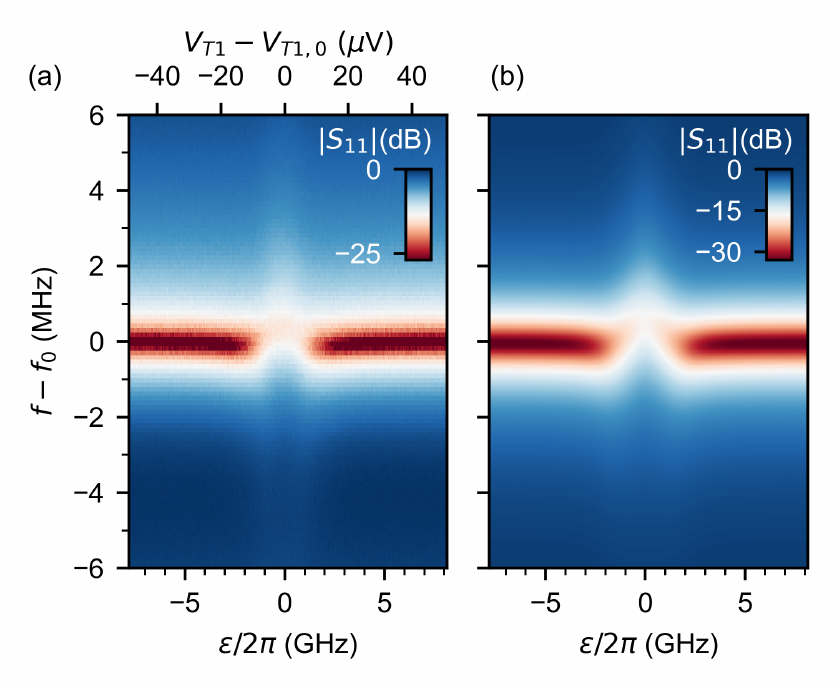}
\caption{\textbf{Response of the resonator in the resonant regime.} (a) Magnitude in dB of the power reflected at the resonator, as $V_\text{T1}$ is varied across the $(6,9)\longleftrightarrow(7,8)$ ICT. (b) Simulation results based on Eq.~\ref{eq1}.}
\label{fig4}
\end{figure}

In our split-gate geometry, we are not able to tune the tunnel coupling independently, however $t_\text{c}$ can vary with electron occupation. We bias the DQD to a different ICT (highlighted by the green dashed rectangle in Fig.~\ref{fig1}(d)) and investigate the effect of the DQD on the resonator in this case. In Fig.~\ref{fig4}(a), we measure the resonator's frequency spectrum as we vary $\varepsilon$ across the $(6,9)\longleftrightarrow(7,8)$ transition, and observe a substantially different response with respect to the dispersive regime (also note the reduced span of the horizontal axis). In this case, close to $\varepsilon=0$, the resonator's frequency increases as opposed to the previous configuration. In addition, we again observe an increase of the effective resonator linewidth $\kappa^*$. We attribute these effects to the qubit and the resonator becoming resonant at some $\varepsilon$ where $\Omega=f_\text{0}$. However, due to the a large DQD decay rate, we are unable to observe the vacuum Rabi mode splitting characteristic of the strong coupling regime. To confirm our hypothesis, we repeat the process described in Appendix~\ref{appendix:sparam} to fit the data. Using the lever arm measurement $\alpha = 0.66 \pm 0.01$, obtained from magnetospectroscopy of this ICT, we find a coupling strength $g_0/(2\pi)=34.77 \pm 0.06$~MHz, tunnel coupling $2t_\text{c}/(2\pi)=1065 \pm 2$~MHz and decay rate $\gamma/(2\pi)=2.334 \pm 0.007$~GHz. The smaller lever arm accounts for part of the reduction in $g_0$ compared with the dispersive regime, and a further reduction may be linked to the smaller tunnel coupling, analogous to the dependence of the coherent coupling rate on Josephson energy for superconducting qubits \cite{koch2007charge}. The reduced $\gamma$ in comparison to the dispersive regime can be understood by assuming phonon-mediated relaxation dominates, for which $\gamma \propto t_\text{c}^3$ \cite{boross2016valley}. Our simulations confirm that the system reaches the resonant regime since the tunnel coupling is lower than the frequency of the resonator. Furthermore, the simulation confirms that the fast decay rate of the DQD prevents us observing the strong coupling limit. 

\section{Conclusion}

We have presented a multi-module integrated circuit for fast readout of the charge state of quantum dot devices enabled by a large charge-photon coupling rate. In our demonstration, the enhanced coupling results from two critical improvements. First, the large interdot gate lever arm which arises from the thin gate oxide, the thin silicon-on-insulator layer, and the asymmetric gate overlap of the nanowire. Second, the enhanced resonator impedance due to the reduced capacitance of the resonator, achieved by coupling inductively rather than capacitively \cite{ahmed2018radio, schaal2020fast, ibberson2019low}. In the dispersive regime, we achieve a frequency shift of the order of the resonator bandwidth, $\chi/\kappa\approx 1$, the condition for maximum state visibility. Furthermore, the measured coupling rate is greater than the resonator linewidth, satisfying one of the requirements for strong coupling. Future experiments aiming to achieve strong coupling to the charge degree of freedom in multi-module quantum systems should focus on decreasing the DQD charge decay rate. One strategy to do this is to move to high-mobility heterostructures where lower $\gamma$ have been demonstrated~\cite{mi2017strong,scarlino2019all}, possibly because of the reduced charge noise when compared to Si/SiO$_2$ interfaces. However, achieving a high $\alpha$ in those systems is a complex technical challenge because of the typically large separation between the gates and the quantum dot. Alternatively, coupling to the spin rather than the charge degree of freedom should provide an optimal route to cQED in hybrid superconducting-CMOS systems, particularly if hole-based devices, which are subject to sizable spin-orbit coupling, are used \cite{crippa2019gate, maurand2016cmos}. Our work shows that, in the short term, it will be possible to reach the large coupling rates necessary for cQED without the necessity to integrate microwave resonators in a CMOS process. In the long-term, asymmetric gate designs made out of high kinetic inductance materials, like TiN, inductively coupled to a single input line, should provide a more scalable route to integrated CMOS cQED. Furthermore, if the large $\alpha$ of CMOS DQDs were combined with even higher impedance resonators, it will be possible to access the ultrastrong coupling regime \cite{forn2019ultrastrong}. We note that to achieve the fast readout presented in this paper, we did not use a JPA showing that there is room for further improvement. Finally, we expect that our approach should impact research aiming at improving the readout fidelity of Majorana-based quantum devices~\cite{de2019rapid, sabonis2019dispersive}.

\section{Acknowledgments}

This research has received funding from the European Union's Horizon 2020 Research and Innovation Programme under Grant Agreement No. 688539 (http://mos-quito.eu). M.F.G.Z. aknowledges support from the Royal Society and the Winton Programme of the Physics of Sustainability. D.J.I is supported by the Bristol Quantum Engineering Centre for Doctoral Training, EPSRC Grant No. EP/L015730/1. T.L. and G.A.O. acknowledge support from EPSRC Cambridge NanoDTC EP/L015978/1 and L.C. acknowledges support from EPSRC Cambridge UP-CDT EP/L016567/1.

\appendix

\renewcommand{\thefigure}{A\arabic{figure}}
\setcounter{figure}{0}

\section{S-parameter analysis and fitting}\label{appendix:sparam}

\begin{figure}
    \centering
    \includegraphics{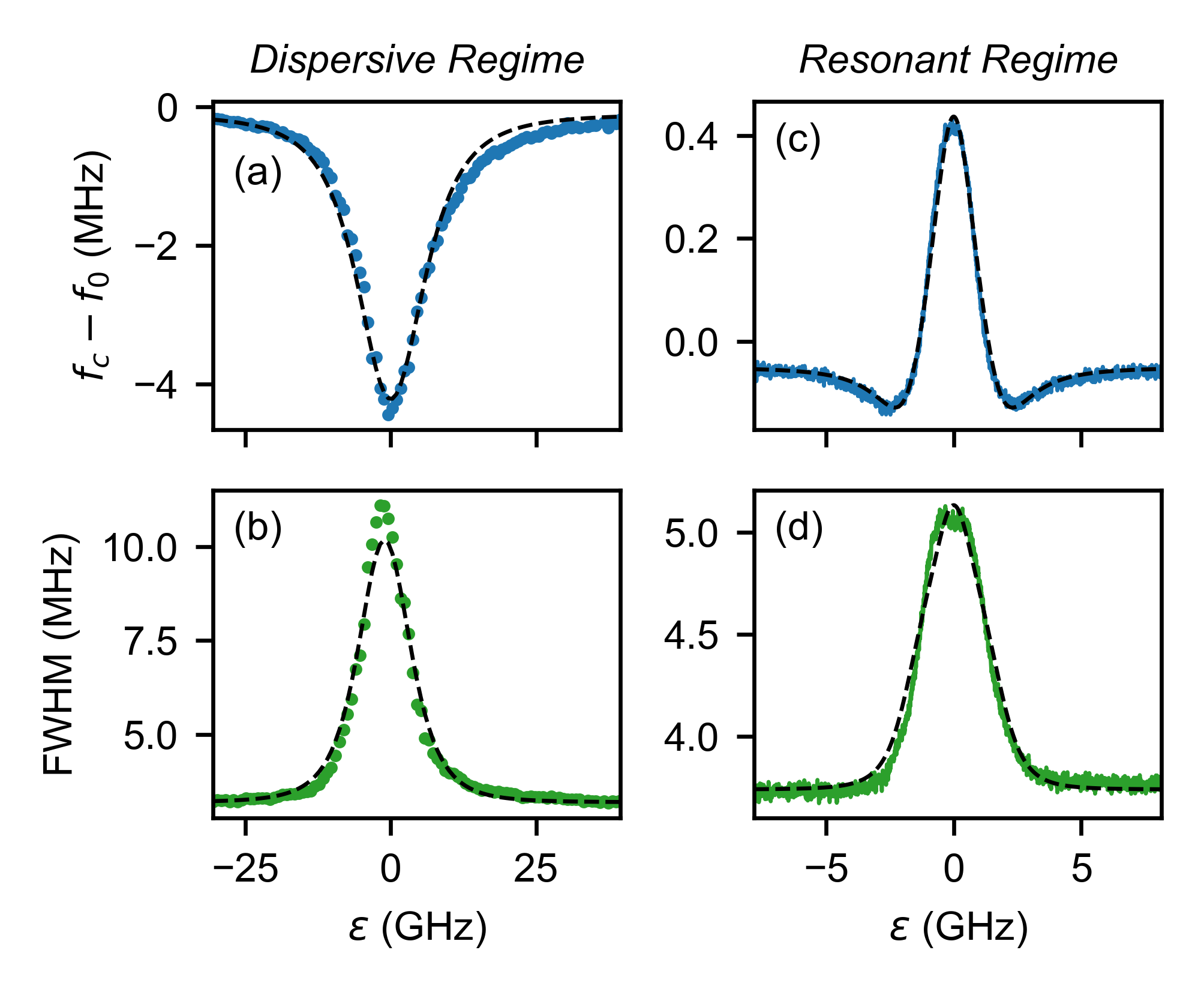}
    \caption{\textbf{Quality of model fitting.} (a) Blue points plot the shift in the centre frequency of the cavity, $f_c$, as the detuning is swept across the ICT in the dispersive regime. (b) Green points plot the change in FWHM of the cavity resonance across the ICT in the dispersive regime. For both (a) and (b) the points are extracted from the data shown in Fig.~\ref{fig2}(c), and the black dashed line represents the best-fit prediction from the model (see Fig.~\ref{fig2}(d)). (c) and (d) plot the same quantities for the ICT in the resonant regime (see Fig.~\ref{fig4}). For all panels errors on $f_\text{c}$ and FWHM are omitted for clarity, and errors on $\varepsilon$ are too small to be visible. In (a) and (b) the $f_\text{c}$ and FWHM error bars are too small to be visible, and for (c) and (d), the errors are roughly the same magnitude as the random scatter in the points.}
    \label{fig:modelfitting}
\end{figure}

For the analysis of the frequency spectrum of the DQD-resonator in Fig.~\ref{fig2} and Fig.~\ref{fig4} in the main paper, it becomes particularly useful to represent Eq.~\ref{eq1} in the form of a complex Lorentzian

\begin{equation}\label{eqA1}
\left|S_{11}\right| = \left| 1 - \frac{i\kappa_\text{ext}}{ \omega-\omega_0+\frac{g_\text{eff}^2\Delta}{\Delta^2+\gamma^2/4} + \frac{i}{2}\left(\kappa+\frac{g_\text{eff}^2\gamma}{\Delta^2+\gamma^2/4}\right) } \right|^2,
\end{equation}

\noindent where $\omega_0 = (2\pi) f_0$ is the resonant angular frequency of the resonator and the center frequency is given by

\begin{equation}\label{eqA2}
\omega_\text{c}=\omega_0-\frac{g_\text{eff}^2\Delta}{\Delta^2+\gamma^2/4}
\end{equation}

\noindent and the effective linewidth of the resonator by 

\begin{equation}\label{eqA3}
\kappa^*=\kappa+\frac{g_\text{eff}^2\gamma}{\Delta^2+\gamma^2/4}.
\end{equation}

In the limit $2t_\text{c}\gg\gamma/2$, Eq.~\ref{eqA2} presents full-width-at-half-maximum in $\varepsilon$ given by $\varepsilon_{1/2}=2\sqrt{(2^{2/3}-1)}t_\text{c}$. 

We now describe our method for extracting the coherent coupling rate $g_0$, tunnel coupling $t_\text{c}$, and charge relaxation rate $\gamma$ from the frequency spectroscopy data, see Fig.~\ref{fig2}(c) for the dispersive regime, and Fig.~\ref{fig4}(a) for the resonant regime. For each step in the detuning $\varepsilon$, we extract the resonance frequency $f_\text{c}$ and the FWHM of the resonance. In Fig.~\ref{fig:modelfitting} we plot $f_\text{c}$ and the FWHM in panels (a) and (b) respectively for the dispersive regime, and (c) and (d) respectively for the resonant regime. We then fit these peak shapes using Eq.~\ref{eqA2} for $f_\text{c}$ and Eq.~\ref{eqA3} for the FWHM. For best accuracy, we fit simultaneously $f_\text{c}$ and FWHM with common parameters $g_0$, $t_\text{c}$ and $\gamma$. Fitting is performed using orthogonal distance regression (ODR), taking into account errors on both $x$ and $y$ axes for each plot. The dominant error on $\varepsilon$ stems from the uncertainty in $\alpha$, and the errors on $f_\text{c}$ and the FWHM are given by the nonlinear least-squares fits used to extract their values from the $\left| S_{11} \right|$ data. The errors quoted on the fitted parameters, $g_0$, $t_\text{c}$, and $\gamma$, are obtained from the covariance matrix of the ODR fit. The quality of the resulting fits, indicated by the black dashed lines, gives high confidence in the calculated parameters.

\section{RF circuit design and operation}\label{appendix:circuit}

\begin{figure}
\centering
\includegraphics[scale=1]{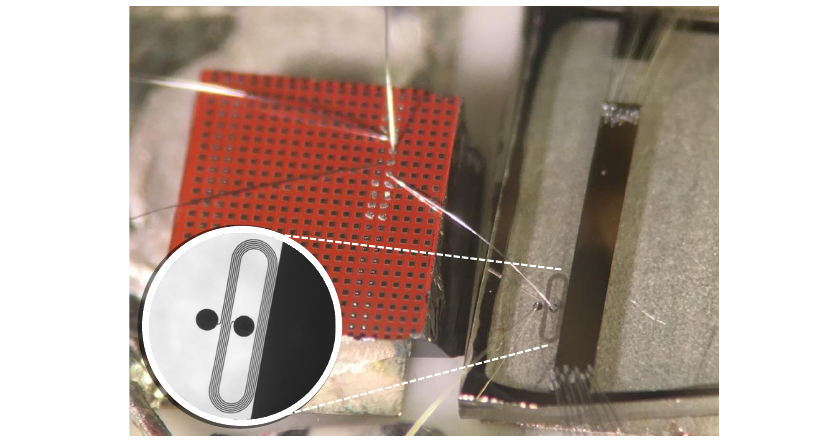}
\caption{\textbf{A photograph of the multi-module set-up.} The silicon chip with an array of square bond-pads is seen to the left, and to the right is the NbN-on-sapphire substrate. The two modules are positioned adjacent to each other on a printed circuit board. The insert magnifies the elongated-spiral inductor adjacent to the superconducting 50$~\Omega$ waveguide.}
\label{Sup_Circuit}
\end{figure}

We present a photograph of the hybrid integrated system described in the main paper, showing the silicon NWFET and superconducting inductor on separate substrates, see Fig.~\ref{Sup_Circuit}. The T1 gate of the silicon CMOS transistor is connected to the centre of the NbN spiral inductor on the adjacent sapphire substrate, using an Al wire of length $\approx 1500~\mu$m and diameter $17.5~\mu$m. An on-PCB bias-tee, not shown, is wire-bonded to the outer bond-pad of the spiral inductor, with an Al wire of similar length. The 50~$\Omega$ microstrip waveguide runs approximately vertically on the right of the figure. The insert magnifies the elongated 6-turn spiral inductor which is separated from the microstrip waveguide by a $4~\mu$m gap; the overall inductor footprint is $812 \times 326~\mu$m, including bond-pads. 

The high impedance of the silicon NWFET split-gate is impedance-matched to the 50~$\Omega$ microstrip waveguide, by embedding the device in an LC resonator and designing the distance between the spiral and the waveguide to achieve critical inductive coupling. The bias-tee capacitor (NPO COG 100~pF high-Q) acts as a short-circuit at the resonant frequency, thereby allowing high quality factors to be achieved by isolating the bias-tee resistor.

\begin{figure}
\centering
\includegraphics[scale=1]{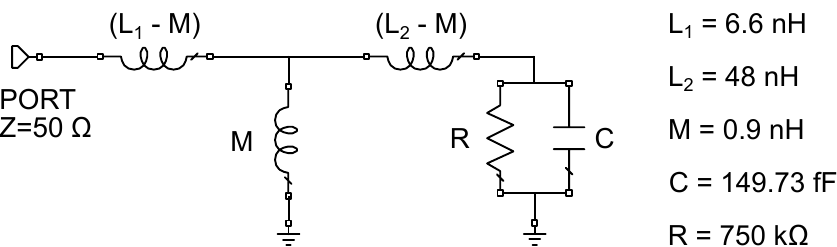}
\caption{\textbf{Simplified circuit model.} The inductively coupled spiral inductor and microstrip waveguide can be approximated as a T-circuit. All of the parasitic capacitances in the system are combined with the gate capacitance into the one capacitor shown. The total losses are accounted for by the 750~k$\Omega$ resistor.}
\label{fig:suppl-Tcircuit}
\end{figure}
 
\begin{figure}
\centering
\includegraphics[scale=1]{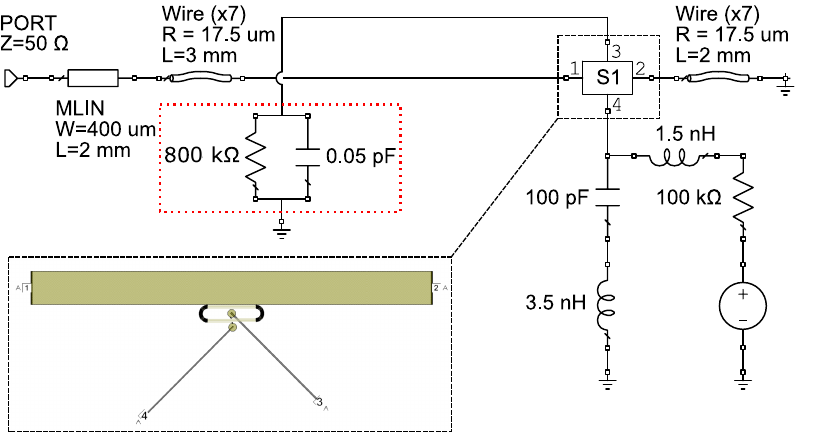}
\caption{\textbf{AWR AXIEM EM simulations.} We model the system, including parasitic contributions, using a 3D planar method-of-moments (MoM) EM analysis simulator. The insert shows the model of the sapphire chip with the microstrip waveguide and spiral inductor. The NWFET is represented by a $800~\text{k}\Omega$ resistor in parallel with a 50~fF capacitor, highlighted by a red dotted box. The MLIN component resides on the Rogers 4003C PCB.}
\label{fig:suppl-AXIEM}
\end{figure}

We now present two models of the resonant system formed by the NWFET, PCB, superconducting chiplet and bond wires, and attempt to reproduce the reflected spectrum data in Fig.~\ref{fig1}(c). The first model consists of lumped elements only, illustrated in Fig.~\ref{fig:suppl-Tcircuit}. In the second model, we use the Cadence\textsuperscript{\textregistered} AWR\textsuperscript{\textregistered} AXIEM\textsuperscript{\textregistered} planar method-of-moments (MoM) software to simulate the NbN/sapphire/grounded-metal structure, which is then embedded in a lumped-element circuit representing the other structures that is simulated using Cadence\textsuperscript{\textregistered} AWR\textsuperscript{\textregistered} Microwave Office\textsuperscript{\textregistered}, see Fig.~\ref{fig:suppl-AXIEM}.

\begin{figure}
\centering
\includegraphics[scale=1]{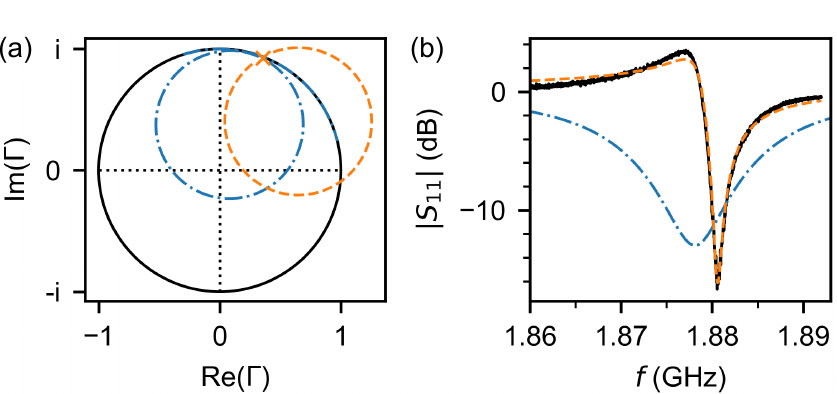}
\caption{\textbf{Simulated resonator response.} (a) Simulated reflection coefficient is plotted on a Smith chart for a range of frequencies around resonance. The orange dashed line gives the response simulated by the simplified circuit model shown in Fig.~\ref{fig:suppl-Tcircuit} which has rotated about its origin on the edge of the circle by $48^{\circ}$ to reproduce the asymmetry observed in the magnitude data. The blue dash-dot line gives the response simulated using the combination shown in Fig.~\ref{fig:suppl-AXIEM}. (b) We plot the absolute value of the reflection coefficient on the dB scale against frequency. The experimental data from Fig.~\ref{fig1}(c) is included in black.}
\label{fig:suppl-simResults}
\end{figure}

In the first model, the coupled inductances of the NbN waveguide and spiral are represented using a T-circuit of inductors (see Fig.~\ref{fig:suppl-Tcircuit}), where the central shunt inductance equals the mutual inductance, $M$, and the left and right inductors are the stray inductances of the waveguide and spiral respectively, where $L_1$ is the inductance of the waveguide and $L_2$ is the spiral inductance. The capacitance, $C$, that completes the resonator includes the self-capacitance of the spiral, the gate capacitance of the NWFET, and the total parasitic capacitances to ground. The resistor, $R$, accounts for internal losses, namely dielectric losses of the two substrates and resistance of non-superconducting metals in the bias tee. We find from MoM simulations of the sapphire chip that $M=0.9$~nH, and deduce from our experimentally measured value of the spiral inductance (equal to $L_2 - M$) that $L_2=48$~nH. The remaining parameters are then obtained by fitting the model to the $\left| S_{11} \right|$ data in Fig.~\ref{fig1}(c) with $M$ and $L_2$ fixed. The asymmetry of the data is accounted for by including an additional parameter in the fit function, specifying the angle of rotation in the complex plane about the point marked by an orange X in Fig.~\ref{fig:suppl-simResults}(a). This rotation is caused experimentally by interference of the radiation reflected from the resonator with radiation reflected at other boundaries, in this case most likely the grounded end of the waveguide \cite{khalil2010loss, sage2011study}. Note that this model does not include the bias tee, making the assumption that it acts as a perfect ground in the vicinity of the resonant frequency.

In the second model, we construct a planar model of the sapphire chip including, from bottom to top, the grounded metal plane on top of the PCB, the sapphire of thickness 500~$\mu$m and dielectric constant 11.58, and the NbN microstrip and spiral. We model the NbN film as a perfect conductor with sheet inductance 1.35~pH/$\square$ to account for kinetic inductance, similar to other estimates \cite{hayashi2014design}. This value was obtained by matching the simulated to the experimental resonant frequency of the un-bonded spiral, which was measured at 3.04 GHz. Wire bonds are also modelled in the MoM simulation in order to connect the spiral to the off-chip elements. Connected to the outer bond-pad via one of these wires is the bias tee, consisting of a 100~pF capacitor to ground, and a 100~k$\Omega$ resistor through which the $V_\text{T1}$ bias is applied. In addition, two inductors of a few nH each are included to model PCB tracks and thru-holes. Connected to the inner bond-pad of the spiral, are a resistor and capacitor in parallel to model the NWFET, similar to the previous model. We find that, to match the experimental resonant frequency, this capacitance must be reduced to $\sim 50$~fF, indicating that the majority of the 150~fF of the lumped-element model originates from the superconducting spiral. Moreover, separate simulations of the un-bonded spiral also show that its capacitance is approximately 100~fF, which could be reduced in future experiments \cite{peruzzo2020surpassing}. We have investigated the effects of different lengths of the bond-wires, which each contribute approximately 2~nH of inductance. For the bond-wire between spiral and NWFET, varying its length $\pm$50$\%$ produces $<$5$\%$ variation in the resonant frequency. A 50\% variation in the length of the bond-wire between the spiral and the bias-tee shifts the resonant frequency by $<$3$\%$. No significant changes to the resonance width or depth were observed.

The simulated reflection spectra of the two models are compared with the experimental data in Fig.~\ref{fig:suppl-simResults}(b). The lumped-element model shows excellent agreement with the data, whereas for the MoM simulation, while its resonant frequency is close to the experiment, the width is overestimated. This can be attributed to our choice of value for the resistor modelling internal losses, for which we have used a value similar to the lumped-element model. This value is clearly an underestimate, from which we can conclude that the internal losses of the spiral predicted by the the MoM simulation have added a significant contribution. Note that varying the bias tee capacitor and resistor values by their rated tolerances has no visible effect. The MoM simulation also does not reproduce the asymmetry of the data.


\section{Frequency multiplexing}\label{appendix:freqmux}

\begin{figure}
\centering
\includegraphics[scale=1]{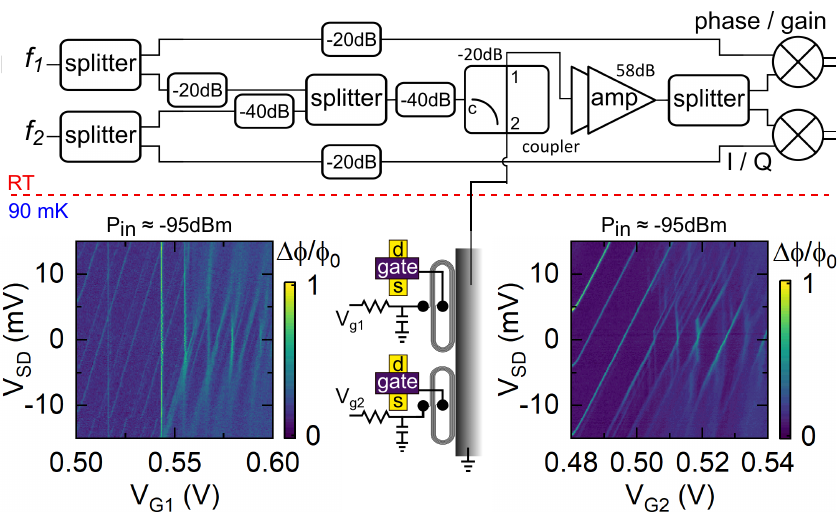}
\caption{\textbf{Schematic of the frequency multiplexing experiment and results.} Tones from separate microwave sources ($f_1$ and $f_2$) are combined and applied to two single-gate silicon NWFETs via a common microwave waveguide. The reflected signal is amplified at room temperature, divided, and then demodulated with the relevant reference signal ($f_1$ is demodulated into phase and gain components using an Analog Devices AD8302, whereas $f_2$ is demodulated into $I$ and $Q$ using a Polyphase AD0540B). Coulomb diamonds for both devices were measured simultaneously at 1.78~GHz (left plot) and 2.45~GHz (right plot).}
\label{fig:freqmux}
\end{figure}

\begin{figure}
\centering
\includegraphics[scale=1]{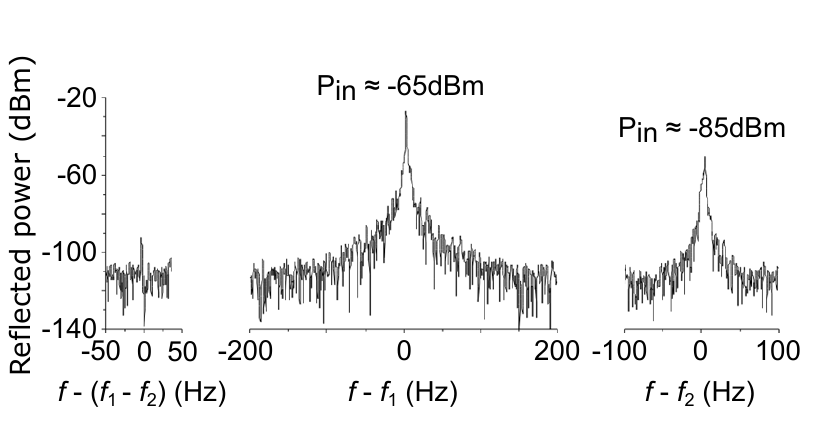}
\caption{\textbf{Measurement of the intermodulation product, $f_1-f_2$, to characterise crosstalk.} The intermodulation product was only observable in the multiplexing experiments when the input power of $f_1$ ($f_2$) was raised by $\sim 30~$dB ($\sim 10~$dB) above the typical experimental level. The resulting product was at least 60 dB (40 dB) below the reflected $f_1$ ($f_2$) level.}
\label{fig:Xtalk}
\end{figure}

We fabricated a second NbN-on-sapphire chip that included multiple inductors close to the microstrip waveguide for frequency multiplexing experiments. Each inductor was wire-bonded to separate bias-tees and single-gate CMOS devices in the same configuration as before, enabling simultaneous measurement of two independent silicon quantum dots via the single microwave line. A 47~nH spiral inductor was bonded to a NWFET with gate length (L) 50~nm and width (W) 30~nm (device 1), and a 29~nH inductor was wire bonded to a NWFET with dimensions 50~nm (L) x 60~nm (W) (device 2), giving resonant frequencies at $f_1 = 1.78~$GHz and $f_2 = 2.45~$GHz respectively. A two-tone probe signal was synthesised by combining $f_1$ and $f_2$ from individual microwave generators and applied to the microstrip waveguide via an unattenuated coaxial line in our dilution refrigerator, see Fig.~\ref{fig:freqmux}. The reflected signal was amplified at room temperature, split into two channels, and then demodulated separately with the generated $f_1$ and $f_2$ tones. By manipulating the gate and bias voltages of the two NWFETs at the same time, we simultaneously acquired the two Coulomb diamond plots shown in Fig.~\ref{fig:freqmux}. We quantify the crosstalk between the two measurements by measuring the signal power of the intermodulation product at $f_1 - f_2$, see Fig.~\ref{fig:Xtalk}. It is not possible to observe the intermodulation product above the noise floor when applying the same input powers used to acquire the Coulomb diamonds, however, increasing the power of the tones resulted in an observable intermodulation product, at least 60 dB (40 dB) below the reflected $f_1$ ($f_2$) tones. The power of the intermodulation product is independent of the devices' gate voltages, indicating that there is no crosstalk at the resonator-device level.

Using separate microwave generators is practical in ``proof-of-principle'' experiments requiring two or three frequencies, however, future experiments should focus on exploiting techniques used in software defined radio, where a multi-frequency signal is synthesised using digital signal processing, before being up-converted to microwave frequencies using a mixer and a reference source \cite{jerger2012frequency}.

\section{Electron population}\label{appendix:elecronpop}

\begin{figure}
    \centering
    \includegraphics{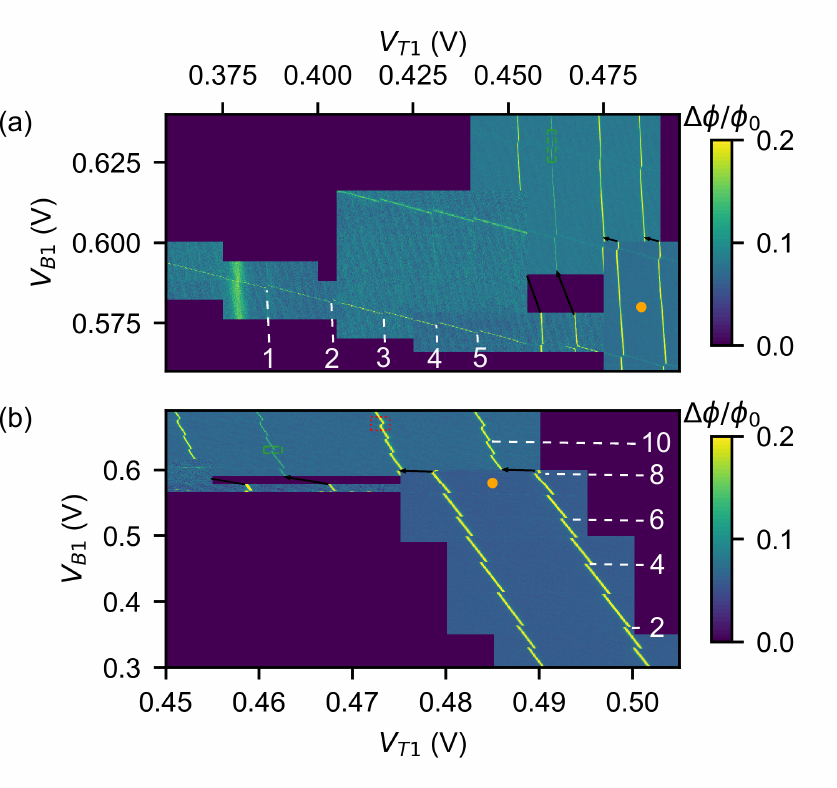}
    \caption{\textbf{Determining the electron occupancy.} (a) Charge stability plot focusing on a charge transition of the B1 dot, where we indicate loading of electrons into the T1 dot with the white dashed lines. (b) Charge stability plot showing a transition of the T1 dot where we indicate loading of electrons into the B1 dot. The red dotted and green-dashed) boxes indicate the ICTs investigated in Figs.~\ref{fig2} and \ref{fig4} respsectively, similar to those in Fig.~\ref{fig1}(d). An orange marker is included in both plots at the same voltage coordinates $(V_\text{T1}, V_\text{B1})=(0.485, 0.58)$~V to aid comparison. Due to an extended time period between some of the measurements, there is a variation in the position of T1-dot charge transitions due to voltage drift; black arrows are used to connect these transitions between the data sets.}
    \label{fig:electronPop}
\end{figure}

Here we present the data used to determine the electron population of the DQD. In Fig.~\ref{fig:electronPop}(a), we zoom in on a charge transition of the B1 dot and count the discontinuities which indicate loading of electrons into the T1 dot, marked by the white dashed lines. No discontinuities are observed at $V_\text{T1} < 0.375$~V, so we conclude that the first shown here, occurring at $V_\text{T1}\approx 0.387$~V, corresponds to the first electron loaded into the T1 dot. In Fig.~\ref{fig:electronPop}(b), we follow a transition of the T1 dot and count electrons loading into the B1 dot. No more discontinuities of this T1 transition are observed for $V_\text{B1}<0.3$~V. The ICT investigated in Fig.~\ref{fig2}(\ref{fig4}) is indicated by the red dotted (green dashed) box, similarly to in Fig.~\ref{fig1}(d).

\section{Large dispersive shift with a different NWFET}

\begin{figure}
    \centering
    \includegraphics{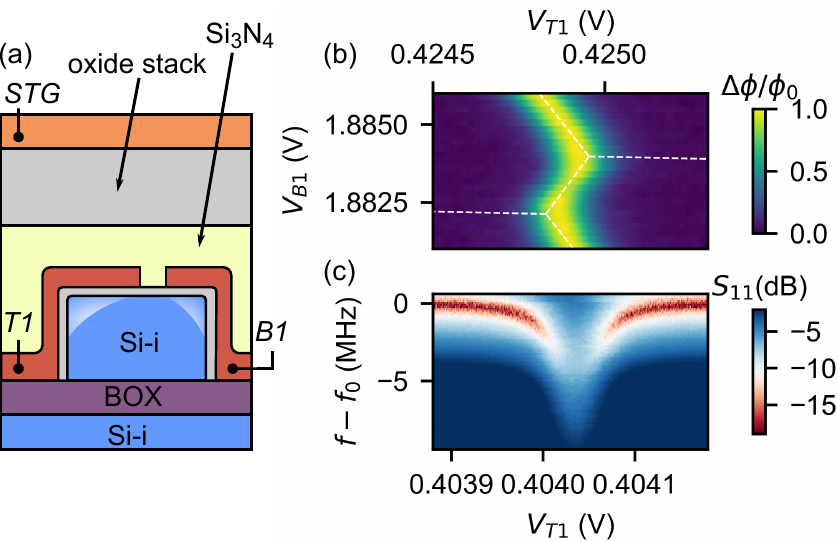}
    \caption{\textbf{Large dispersive interaction in a NWFET with a super-top gate (STG).} (a) Cross-section schematic of the NWFET, similar to Fig.~1(b), but with an additional gate STG above the split-gate. (b) Reduced phase response of the resonator $\Delta \phi/\phi_0$ in the region of an ICT. (c) Magnitude of the signal reflected from the cavity showing the shift in the resonant frequency as $V_\text{T1}$ is swept across the ICT. Note the change in $V_\text{T1}$ axes is due to a change in the $V_\text{STG}$ setting: $V_\text{STG} = 11.15$~V in (a) and 12.1~V in (b).}
    \label{fig:suppl-dev2}
\end{figure}

In this section, we show data obtained using our resonator with a different NWFET device. We exchange the device measured in the main section of the paper for one with a metal super-top gate (STG) which fully covers the poly-Si split gates, as shown in Fig.~\ref{fig:suppl-dev2}(a). The super-top gate is separated from the split-gates by 30~nm of Si$_3$N$_4$ followed by a 350~nm oxide stack. The nanowire and gate dimensions are the same as the previous device, but the devices are on different chips. The cavity resonance decreases slightly to $f_0=1.814$~GHz, which may be caused by a combination of the larger area of this silicon chip, and a small increase in the bond wire length. We tune the voltages of the three gates to the ICT shown in Fig.~\ref{fig:suppl-dev2}(b), where we estimate the total electron occupancy of the DQD to be around 20, similar to the dispersive regime in the previous device. We observe a large dispersive shift of 3.6~MHz in the cavity frequency, shown in Fig.~\ref{fig:suppl-dev2}(c), similar to the shift observed in Fig.~\ref{fig2}(c) with the previous device. These results demonstrate the reproducibility of our approach across different devices.

\section{Condition of maximum state visibility for singlet-triplet readout}\label{appendix:readoutheory}

In order to calculate the condition of maximum state visibility, we calculate the maximum change in reflection coefficient between a singlet outcome, which induces a frequency shift $\chi$, and triplet outcome that does not shift the resonator frequency~\cite{mizuta2017quantum}. Using Eq.~\ref{eq1}, we find:

\begin{align}
\label{eqA4}
\begin{split}
 \Gamma_\text{s}=1+\frac{i\kappa_\text{ext}}{\Delta_0-\chi-i\kappa/2} ,
\\
 \Gamma_\text{t}=1+\frac{i\kappa_\text{ext}}{\Delta_0-i\kappa/2}.
\end{split}
\end{align}

We calculate the change in reflection coefficient $\Delta\Gamma=\Gamma_\text{s}-\Gamma_\text{t}$ and find:

\begin{equation}\label{eqA5}
 \Delta\Gamma=i\frac{\kappa_\text{ext}\chi}{(\Delta_0^2-\Delta_0\chi-\kappa^2/4)-i\kappa(\Delta_0-\chi/2)}.
\end{equation}

The first condition for maximum state visibility occurs when Re$(\Delta\Gamma)=0$. This requirement set the optimal operation detuning from the bare cavity frequency at $\Delta_0=\chi/2$, i.e. at the average frequency between singlet and triplet outcomes. In this case, 

\begin{equation}\label{eqA6}
 \Delta\Gamma=\frac{i4\kappa_\text{ext}\chi}{\chi^2+\kappa^2}.
\end{equation}

Equation~\ref{eqA6} is maximum when $\chi=\kappa$, the condition of maximum state visibility. At this point $\Delta\Gamma=i\frac{2\beta}{1+\beta}$, where $\beta=\kappa_\text{ext}/\kappa_\text{int}$ is the coupling coefficient. Finally, we calculate the signal-to-noise ratio, knowing that photons leak out of the resonator at a rate $\kappa_\text{ext}$

\begin{equation}\label{eqA7}
 \text{SNR}=|\Delta\Gamma|^2\frac{\bar{n}}{n_\text{noise}}\frac{\kappa_\text{ext}}{2\pi}t_\text{int}
\end{equation}

\noindent where $n_\text{noise}=k_\text{B}T_\text{N}/hf_0$ is the photon noise and $\bar{n}$ is the number of photons in the resonator. At the condition for maximum state visibility, for a critically coupled resonator driven at the critical photon number $ \bar{n}=n_\text{crit}=\Delta^2/(4g_0^2)$, we find

\begin{equation}\label{eqA8}
 \text{SNR}=\frac{hf_0}{k_\text{B} T_\text{N}} \frac{g_0^2}{\kappa} t_\text{int}
\end{equation}

\bibliography{references.bib}
\end{document}